\documentclass[reprint,aps,pra,showpacs,floatfix,onecolumn]{revtex4-1}
\usepackage{graphicx}
\usepackage{dsfont}
\usepackage{epstopdf}

\newcommand{\be}{\begin{equation}}
\newcommand{\ee}{\end{equation}}

\begin{document}

\title{Perfect state transfer by means of discrete-time quantum walk on complete bipartite graphs}

\author{M. \v Stefa\v n\'ak\email[correspondence to:]{martin.stefanak@fjfi.cvut.cz}}
\affiliation{Department of Physics, Faculty of Nuclear Sciences and Physical Engineering, Czech Technical University in Prague, B\v
rehov\'a 7, 115 19 Praha 1 - Star\'e M\v{e}sto, Czech Republic}

\author{S. Skoup\'y}
\affiliation{Department of Physics, Faculty of Nuclear Sciences and Physical Engineering, Czech Technical University in Prague, B\v
rehov\'a 7, 115 19 Praha 1 - Star\'e M\v{e}sto, Czech Republic}


\date{\today}

\begin{abstract}
We consider a quantum walk with two marked vertices, sender and receiver, and analyze its application to perfect state transfer on complete bipartite graphs. First, the situation with both the sender and the receiver vertex in the same part of the graph is considered. We show that in this case the dynamics of the quantum walk is independent of the size of the second part and reduces to the one for the star graph where perfect state transfer is achieved. Second, we consider the situation where the sender and the receiver vertex are in the opposite parts of the graph. In such a case the state transfer with unit fidelity is achieved only when the parts have the same size.
\end{abstract}

\maketitle

\section{Introduction}
\label{sec1}

Quantum walks \cite{adz} were proposed as quantum mechanical extensions of classical random walks on a graph or lattice. The time evolution of the quantum walk can be either continuous \cite{fg} or discrete \cite{meyer}. The relation between the continuous-time and discrete-time quantum walks was studied intensively \cite{strauch:rel,childs:rel,yutaka:rel}.

In continuous-time quantum walk \cite{fg,childs:cqw} the evolution of the particle is governed by the Schr\"odinger equation where the Hamiltonian is given by the discrete Laplacian of the graph. Continuous-time quantum walks have found promising applications in quantum search algorithms \cite{childs:search,childs:cont:coin} and modeling of coherent transport on graphs and networks \cite{muelken:dendr,muelken:net,muelken:rev}. Moreover, continuous-time quantum walks were shown to be universal tools for quantum computation \cite{childs}. While the original continuous-time quantum walk search algorithms were analyzed on symmetric lattices or graphs, such as hypercube or a complete graph, later it was found \cite{janmark:search,novo:dim:red,meyer:search} that high symmetry is not required for the optimal runtime of the algorithm. More recently, it was shown \cite{shantanav:search} that the continuous-time quantum walk search algorithm is optimal for almost all graphs.

For discrete-time quantum walks Meyer has shown \cite{meyer} that in order to obtain a non-trivial evolution the system cannot be scalar. A straight-forward way to overcome this constraint is to embed the particle with an additional internal degree of freedom, usually called the coin, which governs the displacements of the particle \cite{ambainis:1d,mackay:2d}. However, several variants of coinless discrete-time quantum walks were proposed. In scattering quantum walks \cite{hillery:interfer,feldman:scattering,feldman:mod}, which were introduced following the analogy with interferometers, the states of the quantum particle corresponds to the directed edges of the graph. Equivalence between the scattering and coined quantum walks was analyzed in detail \cite{equiv:coin:scat,equiv:coin:scat:2}. Szegedy \cite{szegedy} proposed a construction of discrete-time quantum walks based on quantization of classical Markov chains. In the staggered quantum walk model \cite{patel:stagg,patel:stagg:2,falk:stagg,amabainis:stagg,boettcher:stagg,santos:stagg} the evolution of the particle is governed by reflections that correspond to tessellations of the underlying graph. The staggered quantum walk model was recently proven to be more general than both the coined and the Szegedy's walk \cite{portugal:stagg,portugal:stagg:2}. Discrete-time quantum walks were applied to various quantum information tasks including quantum search \cite{szegedy,skw,ambainis,potocek:search,hein:search,reitzner:search,santos} or detecting anomalies in graphs \cite{feldman:struct,hillery:probe,cottrell:find}, and were shown to be universal models of quantum computation \cite{Lovett}.

Quantum walks were also applied to the problem of perfect state transfer \cite{bose:pst} between two vertices of a graph or lattice, which we call sender and receiver. One approach relies on defining the dynamics at each individual vertex in order to achieve state transfer between the sender and the receiver. This method was analyzed in discrete-time quantum walks on a circle \cite{wojcik:qw:pst,gedik:qw:pst} and a square lattice \cite{zhan:qw:pst}. Another possibility is to modify the dynamics only at the sender and the receiver. This method was proposed for wave communication on regular lattices \cite{hein:wave:com} and was further analyzed on various types of finite graphs in \cite{kendon:qw:pst,barr:pst,stef:pst}. Typically, it does not allow for the transfer of the internal state of the particle which is possible in other discrete-time models \cite{wojcik:qw:pst,gedik:qw:pst,zhan:qw:pst}. On the other hand, the method requires less control over the system, since we only have to adjust the coin operators at the sender and the receiver vertices. In the continuous-time quantum walk framework it was shown \cite{shantanav:search} that this protocol achieves perfect state transfer for almost any graph.

In the present paper we extend the results of \cite{stef:pst}, where we have among others considered the perfect state transfer on a star graph by means of discrete-time quantum walk, to complete bipartite graphs. Indeed, star graph is a particular example of a complete bipartite graph where one of the parts has only one vertex. Two different scenarios are considered, namely the sender and the receiver vertex are either located in the same part or in the opposite parts. We show that when both the sender and the receiver vertex are located in the same part the dynamics of the walk is independent of the size of the second part. Hence, the effective evolution operator is the same as for the star graph, where the perfect state transfer is achieved with unit fidelity \cite{stef:pst}. Next, we analyze the situation where the sender and the receiver vertex are in the opposite parts. We show that in such a case the perfect state transfer with unit fidelity is achieved only when the parts are of the same size.

The rest of the paper is organized as follows: In Section~\ref{sec1a} we introduce the notation and review the basic ideas of state transfer by means of discrete-time quantum walk. The case where both the sender and the receiver vertex are located in the same part is considered in Section~\ref{sec2}. In Section~\ref{sec3} the situation where the sender and the receiver vertex are in the opposite parts is analyzed. We summarize our results and present an outlook in the conclusions of Section~\ref{sec4}.

\section{Preliminaries}
\label{sec1a}

In this Section we introduce the notation that will be used later on in the paper. We analyze the state transfer between two vertices of a complete bipartite graph of $m$ plus $n$ vertices $K_{m,n}$ by means of a discrete-time quantum walk. We consider the coined quantum walk model. Alternatively, one can employ the scattering quantum walk formalism \cite{hillery:interfer,feldman:scattering,feldman:mod}. Nevertheless, both models are equivalent \cite{equiv:coin:scat,equiv:coin:scat:2} and lead to exactly the same dynamics. To distinguish between the vertices of different parts of the complete bipartite graph we label them with latin letters in the first part and greek letters in the second.

Let us begin with the definition of the Hilbert space of the quantum walk we consider. Since the graph is bipartite the Hilbert space can be written as a direct sum
$$
\mathcal{H} = \mathcal{H}^{(1)}\oplus\mathcal{H}^{(2)},
$$
where the states from $\mathcal{H}^{(1)}$ ($\mathcal{H}^{(2)}$) corresponds to the particle located in the first (second) part accompanied with some coin state. The Hilbert spaces $\mathcal{H}^{(i)}$ have the form of tensor product
$$
\mathcal{H}^{(i)} = \mathcal{H}^{(i)}_P\otimes\mathcal{H}^{(i)}_C,
$$
of the position space $\mathcal{H}_P^{(i)}$ and the coin space $\mathcal{H}_C^{(i)}$. In the first part, which has $m$ vertices, the position space $\mathcal{H}_P^{(1)}$ is spanned by vectors $|i\rangle_p$ with $i$ ranging from 1 to $m$. The coin space $\mathcal{H}_C^{(1)}$ is determined by the neighboring vertices, i.e. the vertices where the particle can move in a single step. Since we consider complete bipartite graph, the neighbors of any vertex from the first part are all vertices from the second part. Hence, the coin space $\mathcal{H}_C^{(1)}$ is $n$-dimensional and we denote the basis vectors as $|\alpha\rangle_c$ with $\alpha$ ranging from 1 to $n$. To shorten the notation we denote the basis vectors of $\mathcal{H}^{(1)}$ by $|i,\alpha\rangle\equiv|i\rangle_p\otimes|\alpha\rangle_c$. The state $|i,\alpha\rangle$ corresponds to the particle located at the vertex $i$ that will move to the vertex $\alpha$ after the application of the shift operator, which will be defined later. Similarly, for the second part the position space $\mathcal{H}_P^{(2)}$ is spanned by vectors $|\alpha\rangle_p$ with $\alpha$ ranging from 1 to $n$ and the coin space $\mathcal{H}_C^{(2)}$ is spanned by vectors $|i\rangle_c$ with $i$ ranging from 1 to $m$. We denote the basis vectors of $\mathcal{H}^{(2)}$ by $|\alpha,i\rangle\equiv|\alpha\rangle_p\otimes|i\rangle_c$.

The evolution operator which propagates the quantum walk by one step can be decomposed into a sum of two operators
\begin{equation}
\label{evol:op}
U = U_1 + U_2
\end{equation}
where $U_i$ acts nontrivially only on the states from $\mathcal{H}^{(i)}$. The operators $U_i$ have the form
$$
U_i = S_i\cdot C_i,
$$
where $S_i$ denotes the shift operator and $C_i$ is the coin operator. The shift operators displace the particle from one part of the complete bipartite graph to the other according to its coin state. We define them by
\begin{eqnarray}
\nonumber S_1 & = & \sum_{i}\sum_{\alpha} |\alpha,i\rangle\langle i,\alpha|,\\
\nonumber S_2 & = & \sum_{i}\sum_{\alpha} |i,\alpha\rangle\langle \alpha,i| = S_1^\dagger,
\end{eqnarray}
where the summation over $i$ runs from 1 to $m$ and the summation over $\alpha$ runs from 1 to $n$. This will hold throughout the paper unless otherwise specified.

Let us now turn to the coin operators $C_i$ which alter the internal coin states before the shift itself. For the purpose of state transfer we consider two marked vertices, sender and receiver, between which we want to communicate the quantum state. On the marked vertices the coin operator will act in a different way then on the non-marked vertices. In the present paper we consider the coin operator on the non-marked vertices to act as a Grover diffusion operator \cite{grover:search} of appropriate dimension, while on the marked vertices it will act as minus identity. The explicit form of the coin operators $C_1$ and $C_2$ will be given latter depending on the location of the marked vertices. The sender vertex will be always located in the first part and we label it as $s$. The receiver vertex will be either in the same part as the sender vertex, in which case we label it as $r$, or in the second part and we label it as $\rho$.

In the following Sections we analyze the transfer of the particle from the sender vertex to the receiver vertex by means of the discrete-time quantum walk defined above. We start the walk in the initial state
\begin{equation}
\label{init}
|init\rangle = \frac{1}{\sqrt{n}}\sum_{\alpha}|s,\alpha\rangle,
\end{equation}
i.e. the particle is located at the sender vertex with equal-weight superposition of all basis coin states. We analyze its evolution towards the target state, where the particle is located at the receiver vertex. The explicit form of the target state will be given latter depending on the location of the receiver vertex. Our analysis is based on the determination of the invariant subspace of the effective evolution operator which greatly reduces the complexity of the problem. Similar dimensional reduction due to the high symmetry of the graph \cite{krovi} was applied previously  in both discrete-time \cite{feldman:struct,hillery:probe,stef:pst} and continuous-time quantum walks \cite{novo:dim:red}.

\section{Sender and receiver in the same part}
\label{sec2}

Let us begin our analysis with the situation where both the sender and the receiver vertex are in the first part. The coin operator $C_1$ is then given by
$$
C_1 = -(|s\rangle_p\langle s| + |r\rangle_p\langle r|)\otimes I_n + \sum_{i\neq s,r}|i\rangle_p\langle i|\otimes G_n,
$$
where $I_n$ denotes the identity and $G_n$ is the Grover diffusion operator
\begin{equation}
\label{Gn}
G_n = \frac{2}{n}\sum_{\alpha,\beta}|\alpha\rangle_c\langle\beta| - \sum_\alpha |\alpha\rangle_c\langle \alpha|,
\end{equation}
both acting on the $n$-dimensional coin space $\mathcal{H}_C^{(1)}$. Hence, we find that the part of the evolution operator acting on $\mathcal{H}^{(1)}$ reads
\begin{eqnarray}
\nonumber U_1 & = & \frac{2}{n}\sum_{i\neq s,r}\sum_{\alpha,\beta}|\alpha,i\rangle\langle i,\beta| - \sum_i\sum_\alpha|\alpha,i\rangle\langle i,\alpha|.
\end{eqnarray}
On the second part, which does not contain any marked vertices, the coin operator is given by
$$
C_2 = \sum_{\alpha}|\alpha\rangle_p\langle\alpha|\otimes G_m,
$$
where $G_m$ is the Grover diffusion operator on the $m$-dimensional coin space $\mathcal{H}_C^{(2)}$
\begin{equation}
\label{Gm}
G_m = \frac{2}{m}\sum_{i,j}|i\rangle_c\langle j| - \sum_i|i\rangle_c\langle i|.
\end{equation}
The part of the evolution operator acting on $\mathcal{H}^{(2)}$ then reads
\begin{eqnarray}
\nonumber U_2 & = & \frac{2}{m}\sum_{i,j}\sum_\alpha|i,\alpha\rangle\langle\alpha,j| - \sum_{i}\sum_\alpha|i,\alpha\rangle\langle\alpha,i|.
\end{eqnarray}
We start the walk in the state (\ref{init}) and analyze its evolution towards the target state
\begin{equation}
\label{target:same}
|target\rangle = \frac{1}{\sqrt{n}}\sum_{\alpha}|r,\alpha\rangle.
\end{equation}
Note that both states (\ref{init}) and (\ref{target:same}) belong to $\mathcal{H}^{(1)}$. Since the walk is bipartite, it is sufficient to consider only the square of the evolution operator, and in fact we can restrict to the part which acts non-trivially on the initial state. We find that the effective two-step evolution operator is given by
\begin{eqnarray}
\label{Ueff:same}
\nonumber U_{eff} & = & U_2\cdot U_1 = \frac{4}{mn}\sum_i\sum_{j\neq s,r}\sum_{\alpha,\beta}|i,\alpha\rangle\langle j,\beta| - \frac{2}{m}\sum_{i,j}\sum_{\alpha}|i,\alpha\rangle\langle j,\alpha| - \\
& & - \frac{2}{n}\sum_{i\neq s,r}\sum_{\alpha,\beta}|i,\alpha\rangle\langle i,\beta| + \sum_i\sum_\alpha|i,\alpha\rangle\langle i,\alpha|
\end{eqnarray}
In the following we show that the evolution of the quantum walk on the complete bipartite graph $K_{m,n}$ is the same as for the star graph $K_{m,1}$ where the perfect state transfer is achieved \cite{stef:pst}.

First, we determine the invariant subspace of the effective evolution operator (\ref{Ueff:same}) which contains the initial (\ref{init}) and target (\ref{target:same}) states. Clearly, the initial and target states are orthogonal and can be used as the first two basis states of the invariant subspace
$$
|\phi_1\rangle = |init\rangle,\quad |\phi_2\rangle = |target\rangle.
$$
To complete the invariant subspace we choose the last basis state as
$$
|\phi_3\rangle = \frac{1}{\sqrt{n(m-2)}}\sum_{i\neq s,r}\sum_{\alpha}|i,\alpha\rangle.
$$
Simple algebra reveals that the action of the effective evolution operator (\ref{Ueff:same}) on the basis states $|\phi_j\rangle$ is given by
\begin{eqnarray}
\nonumber U_{eff}|\phi_1\rangle & = & \left(1-\frac{2}{m}\right)|\phi_1\rangle - \frac{2}{m}|\phi_2\rangle - \frac{2\sqrt{m-2}}{m}|\phi_3\rangle, \\
\nonumber U_{eff}|\phi_2\rangle & = & -\frac{2}{m}|\phi_1\rangle + \left(1-\frac{2}{m}\right)|\phi_2\rangle - \frac{2\sqrt{m-2}}{m}|\phi_3\rangle, \\
\nonumber U_{eff}|\phi_3\rangle & = & \frac{2\sqrt{m-2}}{m}|\phi_1\rangle + \frac{2\sqrt{m-2}}{m}|\phi_2\rangle + \left(1-\frac{4}{m}\right)|\phi_3\rangle .
\end{eqnarray}
Hence, in the invariant subspace spanned by $|\phi_j\rangle$ the evolution operator (\ref{Ueff:same}) reduces to the following matrix
\begin{equation}
\label{Ueff:same:red}
U_{eff} = \left(
  \begin{array}{ccc}
    1-\frac{2}{m} & -\frac{2}{m} & \frac{2\sqrt{m-2}}{m} \\
    -\frac{2}{m} & 1-\frac{2}{m} & \frac{2\sqrt{m-2}}{m} \\
    -\frac{2\sqrt{m-2}}{m} & -\frac{2\sqrt{m-2}}{m} & 1-\frac{4}{m} \\
  \end{array}
\right),
\end{equation}
which is independent of the size of the second part. In fact, the effective evolution operator (\ref{Ueff:same:red}) is exactly the same as for the star graph $K_{m,1}$ where it was shown \cite{stef:pst} that perfect state transfer is achieved when we choose the number of steps of the quantum walk as the closest even integer to
$$
T = \frac{2\pi}{\arccos\left(\frac{m-4}{m}\right)}.
$$

\section{Sender and receiver in opposite parts}
\label{sec3}

Let us now turn to the case where the sender and the receiver vertices are in the opposite part. To be specific, we consider the sender vertex (labeled $s$) located in the first part and the receiver vertex (labeled $\rho$) in the second part. The coin operator $C_1$ is given by
$$
C_1 = -|s\rangle_p\langle s|\otimes I_n + \sum_{i\neq s}|i\rangle_p\langle i|\otimes G_n,
$$
where $G_n$ is given in (\ref{Gn}). The part of the evolution operator acting on $\mathcal{H}^{(1)}$ then reads
$$
U_1 = \frac{2}{n}\sum_{i\neq s}\sum_{\alpha,\beta}|\alpha,i\rangle\langle i,\beta| - \sum_i\sum_\alpha|\alpha,i\rangle\langle i,\alpha|.
$$
Similarly, the coin operator $C_2$ reads
$$
C_2 = -|\rho\rangle_p\langle\rho|\otimes I_m + \sum_\alpha |\alpha\rangle_p\langle \alpha|\otimes G_m,
$$
where $I_m$ denotes the identity on the $m$-dimensional coin space $\mathcal{H}_C^{(2)}$ and $G_m$ is given in (\ref{Gm}). The part of the evolution operator acting on $\mathcal{H}^{(2)}$ then reads
$$
U_2 = \frac{2}{m}\sum_{i,j}\sum_{\alpha\neq\rho}|i,\alpha\rangle\langle\alpha,j| - \sum_{i}\sum_\alpha|i,\alpha\rangle\langle\alpha,i|.
$$
We again begin the walk in the initial state (\ref{init}) and analyze its evolution towards the target state, which now reads
\begin{equation}
\label{target:opp}
|target_2\rangle = \frac{1}{\sqrt{m}}\sum_{i}|\rho,i\rangle.
\end{equation}
Since the sender and the receiver vertices are located in the opposite parts of the graph the initial state (\ref{init}) and the target state (\ref{target:opp}) do not belong to the same subspace of $\mathcal{H}$. Hence, it is suitable first to apply the evolution operator of the walk (\ref{evol:op}) once on the initial state (\ref{init}). The state of the walk after one step is given by
\begin{equation}
\label{init:opp}
|init_2\rangle = U|init\rangle =  -\frac{1}{\sqrt{n}}\sum_{\alpha}|\alpha,s\rangle,
\end{equation}
which belongs to $\mathcal{H}^{(2)}$, i.e. the same subspace as the target state (\ref{target:opp}). From now on we can again employ the bipartitness of the walk and consider the effective two-step evolution operator
\begin{eqnarray}
\label{Ueff:opp}
\nonumber U_{eff} & = & U_1\cdot U_2 = \frac{4}{mn}\sum_{i\neq s}\sum_j\sum_{\alpha\neq\rho}\sum_\beta |\beta,i\rangle\langle\alpha,j| - \frac{2}{n}\sum_{i\neq s}\sum_{\alpha,\beta}|\beta,i\rangle\langle\alpha,i| - \\
& & - \frac{2}{m}\sum_{i,j}\sum_{\alpha\neq\rho}|\alpha,i\rangle\langle\alpha,j| + \sum_i\sum_\alpha|\alpha,i\rangle\langle\alpha,i|.
\end{eqnarray}
In the following we analyze how close we can get from the state (\ref{init:opp}) towards the target state (\ref{target:opp}) by successive applications of the effective evolution operator (\ref{Ueff:opp}).

We begin with the determination of the invariant subspace of (\ref{Ueff:opp}) which includes the vectors (\ref{target:opp}) and (\ref{init:opp}). In contrast to the previous Section, these two vectors are no longer orthogonal and therefore they cannot be directly used as basis vectors of the invariant subspace. One possibility to choose the basis of the invariant subspace is given by \cite{note}
\begin{eqnarray}
\label{basis:opp}
\nonumber |\phi_1\rangle & = & |\rho,s\rangle, \\
\nonumber |\phi_2\rangle & = & \frac{1}{\sqrt{m-1}}\sum_{i\neq s}|\rho,i\rangle, \\
\nonumber |\phi_3\rangle & = & \frac{1}{\sqrt{n-1}}\sum_{\alpha\neq \rho}|\alpha,s\rangle , \\ |\phi_4\rangle & = & \frac{1}{\sqrt{(m-1)(n-1)}}\sum_{i\neq s}\sum_{\alpha\neq\rho}|\alpha,i\rangle .
\end{eqnarray}
Clearly we find
\begin{eqnarray}
\nonumber |init_2\rangle & = & -\frac{1}{\sqrt{n}}|\phi_1\rangle - \sqrt{\frac{n-1}{n}}|\phi_3\rangle , \\
\nonumber |target_2\rangle & = & \frac{1}{\sqrt{m}}|\phi_1\rangle + \sqrt{\frac{m-1}{m}}|\phi_2\rangle ,
\end{eqnarray}
i.e. both the target state (\ref{target:opp}) and the state (\ref{init:opp}) are included in the subspace spanned by $|\phi_j\rangle$. Simple algebra reveals that the action of the effective two-step evolution operator (\ref{Ueff:opp}) on the basis states (\ref{basis:opp}) is given by
\begin{eqnarray}
\nonumber U_{eff} |\phi_1\rangle & = & |\phi_1\rangle, \\
\nonumber U_{eff} |\phi_2\rangle & = & \left(1-\frac{2}{n}\right)|\phi_2\rangle - 2\frac{\sqrt{n-1}}{n}|\phi_4\rangle ,\\
\nonumber U_{eff} |\phi_3\rangle & = & 4\frac{\sqrt{(m-1)(n-1)}}{m n}|\phi_2\rangle + \left(1-\frac{2}{m}\right)|\phi_3\rangle + \\
\nonumber & & + 2\frac{(n-2)\sqrt{m-1}}{m n}|\phi_4\rangle ,\\
\nonumber U_{eff} |\phi_4\rangle & = & 2\frac{(m-2)\sqrt{n-1}}{m n}|\phi_2\rangle - 2\frac{\sqrt{m-1}}{m}|\phi_3\rangle + \\
\nonumber & & + \frac{(m-2)(n-2)}{m n}|\phi_4\rangle .
\end{eqnarray}
Hence, the effective evolution operator (\ref{Ueff:opp}) is in the invariant subspace spanned by the vectors $|\phi_j\rangle$ given by the matrix
\begin{equation}
\label{Ueff:opp:red}
U_{eff} = \left(
            \begin{array}{cccc}
              1 & 0 & 0 & 0 \\
              0 & 1-\frac{2}{n} & 4\frac{\sqrt{(m-1)(n-1)}}{m n} & 2\frac{(m-2)\sqrt{n-1}}{m n} \\
              0 & 0 & 1-\frac{2}{m} & - 2\frac{\sqrt{m-1}}{m} \\
              0 & - 2\frac{\sqrt{n-1}}{n} & 2\frac{(n-2)\sqrt{m-1}}{m n} & \frac{(m-2)(n-2)}{m n} \\
            \end{array}
          \right).
\end{equation}

Let us now determine the eigenvalues and eigenvectors of the matrix (\ref{Ueff:opp:red}). We find that the eigenvalues are $1$ which has a two-fold degeneracy and $e^{\pm i \omega}$, where the phase $\omega$ is given by
$$
\omega = \arccos\left(\frac{m n - 2m - 2n + 2}{m n}\right).
$$
The eigenvectors corresponding to eigenvalue 1 are given by
\begin{eqnarray}
\nonumber |\chi_1\rangle & = & |\phi_1\rangle, \\
\nonumber |\chi_2\rangle & = & \frac{1}{\sqrt{m+n-1}}\left(\sqrt{n-1}|\phi_2\rangle + \sqrt{m-1}|\phi_3\rangle - |\phi_4\rangle\right) .
\end{eqnarray}
Eigenvectors corresponding to $\lambda_{3,4} = e^{\pm i \omega}$ are given by
\begin{eqnarray}
\nonumber |\chi_3\rangle & = & a|\phi_2\rangle + b|\phi_3\rangle + c |\phi_4\rangle,\\
\nonumber |\chi_4\rangle & = & \overline{a}|\phi_2\rangle + b|\phi_3\rangle + \overline{c} |\phi_4\rangle,
\end{eqnarray}
where the coefficients $a$, $b$ and $c$ read
\begin{eqnarray}
\nonumber a & = & - \frac{m n - m - n + 1}{\sqrt{2n(m-1)(n-1)(m+n-1)}} - \frac{i}{\sqrt{2 n}}, \\
\nonumber b & = & \sqrt{\frac{n}{2(m+n-1)}} ,\\
\nonumber c & = & \frac{m-1}{\sqrt{2n(m-1)(m+n-1)}} - i \sqrt{\frac{n-1}{2n}} .
\end{eqnarray}

Let us now analyze the evolution of the state (\ref{init:opp}) towards the target state (\ref{target:opp}) under the effective evolution operator (\ref{Ueff:opp:red}). The decomposition of the state (\ref{init:opp}) into the eigenbasis $|\chi_i\rangle$ is given by
\begin{eqnarray}
\nonumber |init_2\rangle  & = & -\frac{1}{\sqrt{n}}|\chi_1\rangle - \sqrt{\frac{m n - m - n + 1}{n(m+n-1)}}|\chi_2\rangle - \\
 \nonumber & & - \sqrt{\frac{n-1}{2(m+n-1)}}\left(|\chi_3\rangle +|\chi_4\rangle\right).
\end{eqnarray}
Hence, the state of the walk after $t$ iterations of the effective two-step evolution operator, i.e. after $2t+1$ steps of the quantum walk, reads
\begin{eqnarray}
\label{time:evol:opp}
\nonumber |\psi(2t+1)\rangle & = & -\frac{1}{\sqrt{n}}|\chi_1\rangle - \sqrt{\frac{m n - m - n + 1}{n(m+n-1)}}|\chi_2\rangle - \\
 & & - \sqrt{\frac{n-1}{2(m+n-1)}}\left(e^{i \omega t}|\chi_3\rangle + e^{-i \omega t}|\chi_4\rangle\right).
\end{eqnarray}
We find that the decomposition of the target state (\ref{target:opp}) into the eigenbasis $|\chi_i\rangle$ is given by
\begin{eqnarray}
\nonumber |target_2\rangle  & = & \frac{1}{\sqrt{m}}|\chi_1\rangle + \sqrt{\frac{(m-1)(n-1)}{m(m+n-1)}}|\chi_2\rangle + \\
\nonumber & & + \sqrt{\frac{m-1}{m}}\left(\overline{a}|\chi_3\rangle + a |\chi_4\rangle\right) .
\end{eqnarray}
Therefore, the fidelity between the state of the particle after $2t+1$ steps of the quantum walk (\ref{time:evol:opp}) and the desired target state equals
\begin{eqnarray}
\label{fidel}
\nonumber {\cal F}(2t+1) & = & |\langle\psi(2t+1)|target_2\rangle|^2 \\
\nonumber & = & \frac{1}{mn(m+n-1)^2} \left[mn - (m-1)(n-1)\cos{(\omega t)} + \right.\\
 & & \left. + \sqrt{(m-1)(n-1)(m+n-1)}\sin{(\omega t)}\right]^2.
\end{eqnarray}
Note that for even number of steps the particle is located in the first part of the graph and hence the fidelity vanishes. We find that the first maximum of the fidelity is reached for
$$
\omega t = \arccos\left(-\sqrt{\frac{(m-1)(n-1)}{mn}}\right),
$$
i.e. in order to achieve the state transfer with highest possible probability we have to choose the number of steps of the quantum walk equal to odd integer closest to
\begin{equation}
\label{T:opp}
T = 2t+1 = \frac{2\arccos\left(-\sqrt{\frac{(m-1)(n-1)}{mn}}\right)}{\arccos\left(\frac{m n - 2m - 2n + 2}{m n}\right)} + 1.
\end{equation}
The maximal value of fidelity is given by
\begin{equation}
\label{Fmax}
{\cal F}_{\rm max} = \left(\frac{\sqrt{(m-1)(n-1)}+\sqrt{mn}}{m+n-1}\right)^2.
\end{equation}
We note that the maximal fidelity is less than one unless $m=n$, i.e. perfect state transfer with unit probability can be achieved only when both parts have the same number of vertices.

\begin{figure}[h]
\begin{center}
\includegraphics[width=0.75\textwidth]{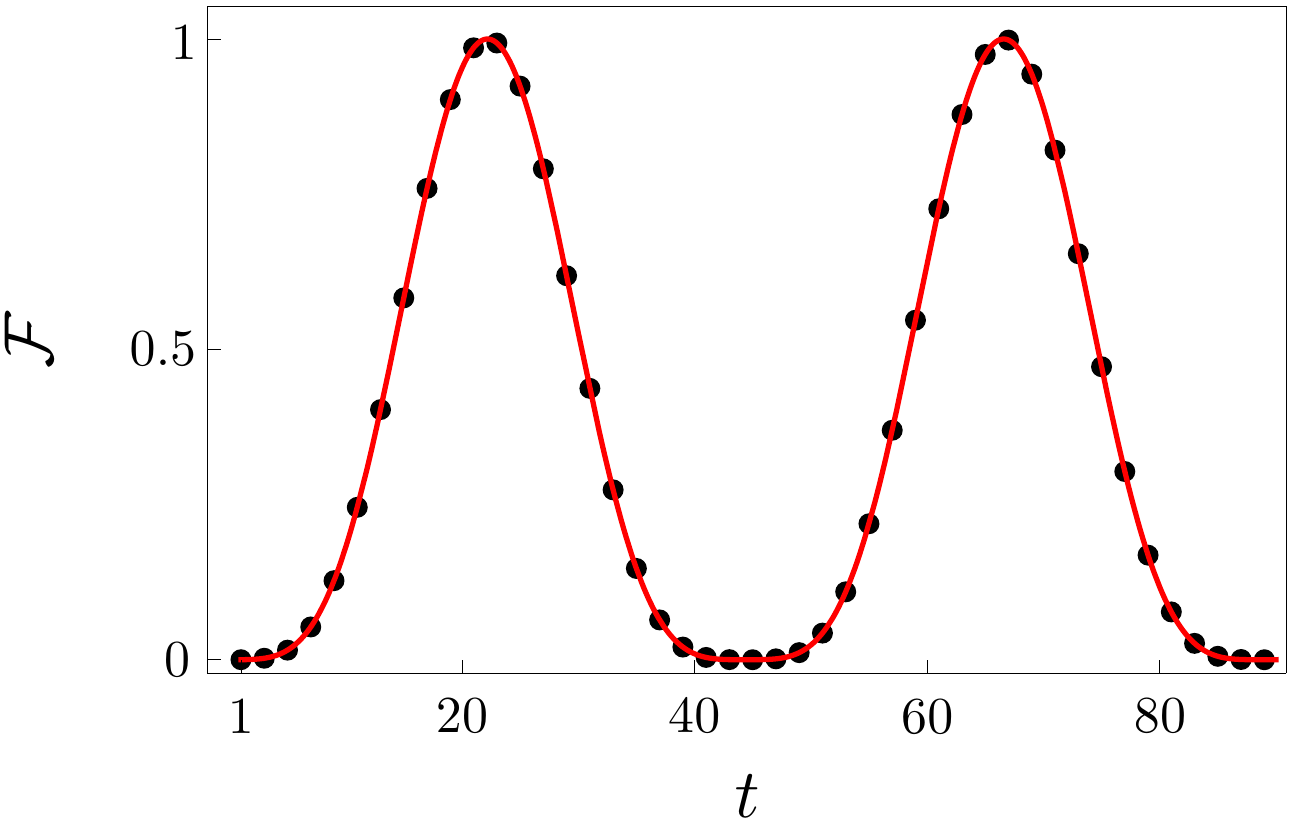}\hfill
\end{center}
\caption{Fidelity as a function of the number of steps $t$ for state transfer on the complete bipartite graph $K_{100,100}$. The red line denotes the analytical expression (\ref{fidel}) and the black dots corresponds to the numerical simulation. The black dots are plotted only for odd time steps, since for even steps the fidelity vanishes. Since $m=n$  perfect state transfer with unit fidelity is possible. The number of steps required to achieve unit fidelity is given by $T=23$ steps, in accordance with the analytical results of (\ref{T:opp}).}
\label{fig1}
\end{figure}

\begin{figure}[h]
\begin{center}
\includegraphics[width=0.75\textwidth]{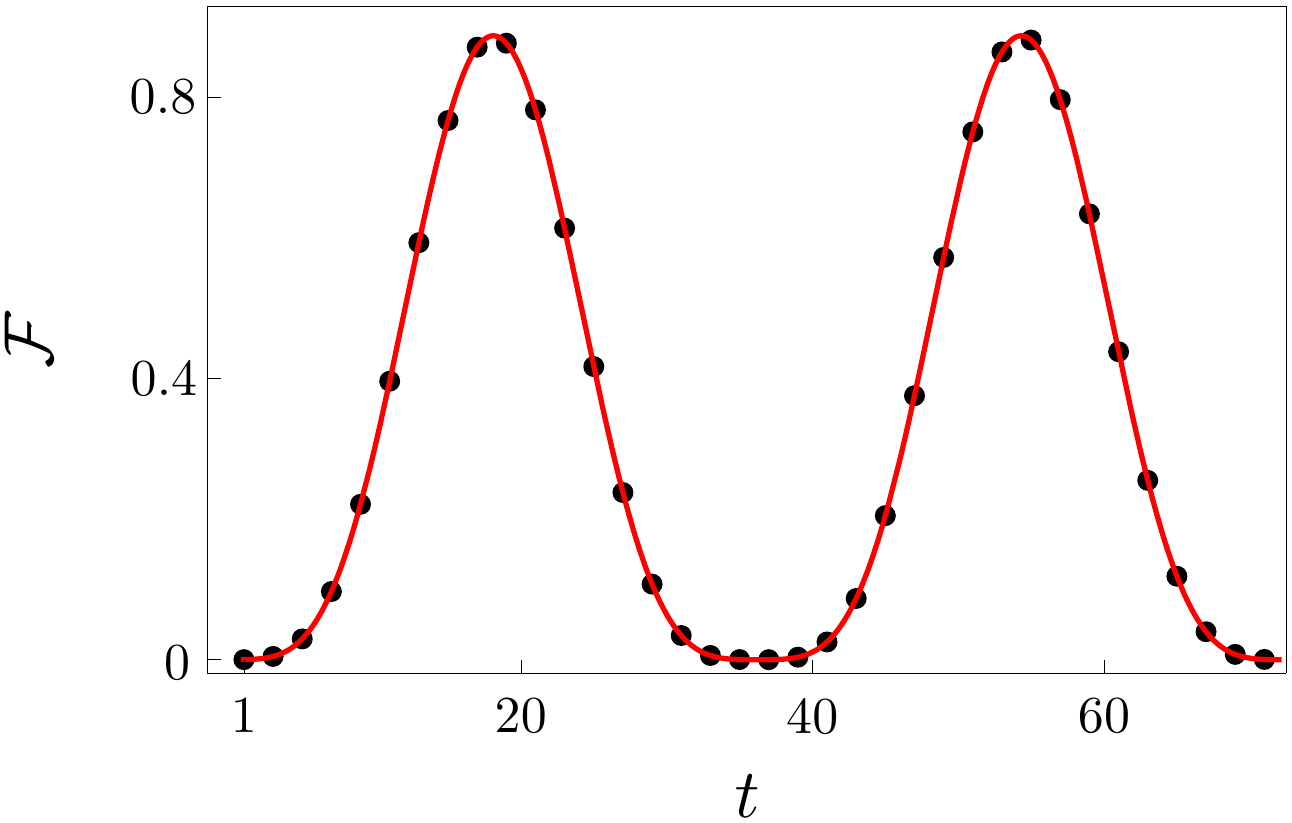}\hfill
\end{center}
\caption{Fidelity as a function of the number of steps for state transfer on the complete bipartite graph $K_{100,50}$. The red line denotes the analytical expression (\ref{fidel}) and the black dots corresponds to the numerical simulation. Fidelity reaches the maximal value of ${\cal F}_{\rm max} \approx 0.89$ after $T=19$ steps of the walk, in accordance with the analytical results of (\ref{Fmax}) and (\ref{T:opp}).}
\label{fig2}
\end{figure}

\begin{figure}[h]
\begin{center}
\includegraphics[width=0.75\textwidth]{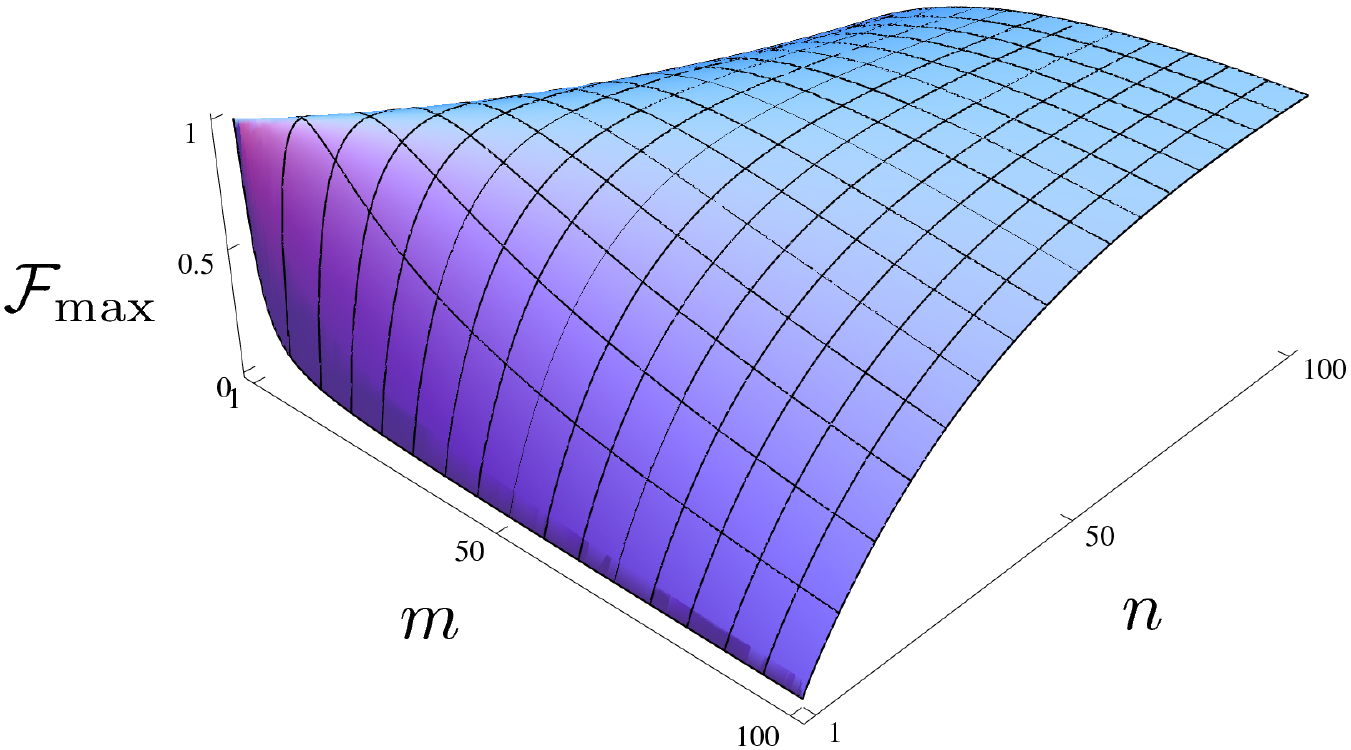}\hfill
\end{center}
\caption{Maximal value of the fidelity (\ref{Fmax}) in dependence on the size of the parts $m$ and $n$ of the complete bipartite graph $K_{m,n}$.}
\label{fig3}
\end{figure}

\begin{figure}[h]
\begin{center}
\includegraphics[width=0.75\textwidth]{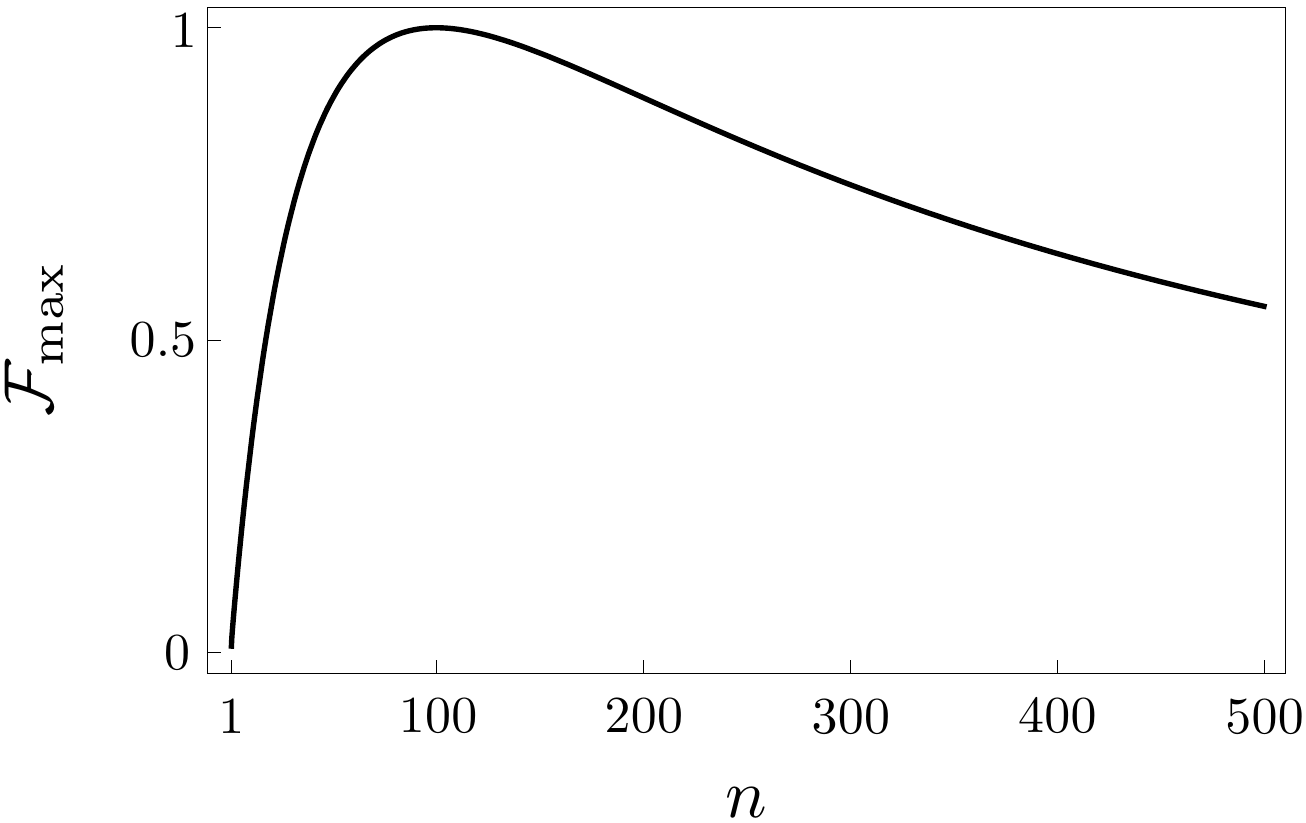}\hfill
\end{center}
\caption{Maximal value of fidelity (\ref{Fmax}) in dependence on the size of the part which contains the receiver vertex $n$. The size of the part which contains the sender vertex is taken as $m=100$. We see that ${\cal F}_{\rm max}$ reaches unity for $m=n=100$ and then declines gradually.}
\label{fig4}
\end{figure}

We illustrate these results in Figures~\ref{fig1}-\ref{fig4}. In Figure~\ref{fig1} we consider state transfer on the complete bipartite graph $K_{100,100}$. The plot shows the fidelity as a function of the number of steps. Since the parts of the graph have the same number of vertices it is possible to achieve perfect state transfer with unit fidelity.

In Figure~\ref{fig2} we show the fidelity as a function of the number of steps for state transfer on the complete bipartite graph $K_{100,50}$, i.e. the parts have different number of vertices. In such a case it is not possible to achieve state transfer with unit fidelity. According to (\ref{Fmax}) the maximal attainable value of fidelity is ${\cal F}_{\rm max} \approx 0.89$.

In Figure~\ref{fig3} we display the maximal value of fidelity (\ref{Fmax}) as a function of the size of the parts $m$ and $n$ of the complete bipartite graph $K_{m,n}$.

Finally, in Figure~\ref{fig4} we show the maximal value of fidelity for the complete bipartite graph $K_{100,n}$ as a function of the size of the part containing the receiver vertex $n$.

\section{Conclusions}
\label{sec4}

State transfer between two vertices of a complete bipartite graph by means of discrete-time quantum walk was analyzed. We have shown that when the sender and the receiver vertices are located in the same part the perfect state transfer is achievable independent of the size of the second component. However, when the sender and the receiver vertices are located in the opposite parts of the graph the state transfer with unit fidelity is achieved only when the two parts of the graph have exactly the same number of vertices. The maximal value of fidelity of state transfer in dependence on the number of vertices of the respective parts of the complete bipartite graph was determined.

In the present model we have discussed state transfer in a closed system, where the dynamics is purely unitary. However, one can consider interactions of the particle with environment, i.e. open quantum dynamics which inevitably leads to decoherence. It would be interesting to analyze how strong interaction one can tolerate in order to achieve state transfer with a desired fidelity. We plan to investigate the effects of decoherence on the state transfer efficiency in the near future.

\begin{acknowledgements}

We acknowledge the financial support from RVO~14000 and from Czech Technical University in Prague under Grant No. SGS16/241/OHK4/3T/14. M\v S is grateful for the financial support from GA\v CR under Grant No. 14-02901P.

\end{acknowledgements}


\begin{thebibliography}{100}

\bibitem{adz}
Y. Aharonov, L. Davidovich and N. Zagury, Quantum random walks, Phys. Rev. A {\bf 48}, 1687 (1993)

\bibitem{fg}
E. Farhi and S. Gutmann, Quantum computation and decision trees, Phys. Rev. A {\bf 58}, 915 (1998)

\bibitem{meyer}
D. Meyer, From quantum cellular automata to quantum lattice gases, J. Stat. Phys. {\bf 85},  551, (1996)

\bibitem{strauch:rel}
F. W. Strauch, Connecting the discrete-and continuous-time quantum walks, Phys. Rev. A 74, 030301 (2006)

\bibitem{childs:rel}
A. M. Childs, On the relationship between continuous- and discrete-time quantum walk, Comm. Math. Phys. {\bf 294}, 581 (2010)

\bibitem{yutaka:rel}
Y. Shikano, From discrete time quantum walk to continuous time quantum walk in limit distribution, J. Comput. Theor. Nanosci. {\bf 10}, 1558 (2013)

\bibitem{childs:cqw}
A. M. Childs, E. Farhi and S. Gutmann, An example of the difference between quantum and classical random walks, Quantum Inf. Process. {\bf 1}, 35 (2002)

\bibitem{childs:search}
A. M. Childs and J. Goldstone, Spatial search by quantum walk, Phys. Rev. A {\bf 70}, 022314 (2004).

\bibitem{childs:cont:coin}
A. M. Childs and J. Goldstone, Spatial search and the Dirac equation, Phys. Rev. A {\bf 70}, 042312 (2004)

\bibitem{muelken:dendr}
O. M\"ulken, V. Bierbaum and A. Blumen, Coherent exciton transport in dendrimers and continuous-time quantum walks, J. Chem. Phys. {\bf 124}, 124905 (2006)

\bibitem{muelken:net}
O. M\"ulken, V. Pernice and A. Blumen, Quantum transport on small-world networks: A continuous-time quantum walk approach, Phys. Rev. E {\bf 76}, 051125 (2007)

\bibitem{muelken:rev}
O. M\"ulken and A. Blumen, Continuous-time quantum walks: Models for coherent transport on complex networks, Phys. Rep. {\bf 502}, 37 (2011)

\bibitem{childs}
A. M. Childs, Universal Computation by Quantum Walk, Phys. Rev. Lett. {\bf 102}, 180501 (2009)

\bibitem{janmark:search}
J. Janmark, D. A. Meyer and T. G. Wong, Global Symmetry is Unnecessary for Fast Quantum Search, Phys. Rev. Lett. {\bf 112}, 210502 (2014)

\bibitem{novo:dim:red}
L. Novo, S. Chakraborty, M. Mohseni, H. Neven and Y. Omar, Systematic Dimensionality Reduction for Quantum Walks: Optimal Spatial Search and Transport on Non-Regular Graphs, Scientific Rep. {\bf 5}, 13304 (2014)

\bibitem{meyer:search}
D. A. Meyer and T. G. Wong, Connectivity is a Poor Indicator of Fast Quantum Search, Phys. Rev. Lett. {\bf 114}, 110503 (2015)

\bibitem{shantanav:search}
S.  Chakraborty, L. Novo, A. Ambainis and Y. Omar, Spatial Search by Quantum Walk is Optimal for Almost all Graphs, Phys. Rev. Lett. {\bf 116}, 100501 (2016)

\bibitem{ambainis:1d}
A. Ambainis, E. Bach, A. Nayak, A. Vishwanath and J. Watrous, One-dimensional quantum walks, Proc. 33rd Ann. Symp. on Theory of Computing (New York: ACM), 37 (2001)

\bibitem{mackay:2d}
T. D. Mackay, S. D. Bartlett, L. T. Stephenson and B. C. Sanders, Quantum walks in higher dimensions, J. Phys. A {\bf 35}, 2745 (2002)

\bibitem{hillery:interfer}
M. Hillery, J. Bergou and E. Feldman, Quantum walks based on an interferometric analogy, Phys. Rev. A {\bf 68}, 032314 (2003)

\bibitem{feldman:scattering}
E. Feldman and M. Hillery, Scattering theory and discrete-time quantum walks, Phys. Lett. A {\bf 324}, 277 (2004)

\bibitem{feldman:mod}
E. Feldman and M. Hillery, Modifying quantum walks: a scattering theory approach, J. Phys. A {\bf 40}, 11343 (2007)

\bibitem{equiv:coin:scat}
F. M. Andrade and M. G. E. da Luz, Equivalence between discrete quantum walk models in arbitrary topologies, Phys. Rev. A {\bf 80}, 052301 (2009)

\bibitem{equiv:coin:scat:2}
B. F. Venancio, F. M. Andrade and M. G. E. da Luz, Unveiling and exemplifying the unitary equivalence of discrete time quantum walk models, J. Phys. A {\bf 46}, 165302 (2013)

\bibitem{szegedy}
M. Szegedy, Quantum Speed-Up of Markov Chain Based Algorithms, in {\it Proceedings of the 45th Symposium on Foundations of Computer Science}, 32–41 (2004)

\bibitem{patel:stagg}
A. Patel, K. S. Raghunathan and P. Rungta, Quantum random walks do not need a coin toss, Phys. Rev. A {\bf 71}, 032347 (2005)

\bibitem{patel:stagg:2}
A. Patel, K. S. Raghunathan and Md. A. Rahaman, Search on a hypercubic lattice using a quantum random walk. II. d=2, Phys. Rev. A 82, 032331 (2010)

\bibitem{falk:stagg}
M. Falk, Quantum search on the spatial grid. arXiv:1303.4127, (2013)

\bibitem{amabainis:stagg}
A. Ambainis, R. Portugal and N. Nahimov, Spatial search on grids with minimum memory, Quantum Inf. Comput. {\bf 15}, 1233 (2015)

\bibitem{boettcher:stagg}
R. Portugal, S. Boettcher and S. Falkner, One-dimensional coinless quantum walks, Phys. Rev. A {\bf 91}, 052319 (2015)

\bibitem{santos:stagg}
R. A. M. Santos, R. Portugal and S. Boettcher, Moments of coinless quantum walks on lattices, Quantum Inf. Process. {\bf 14}, 3179 (2015)

\bibitem{portugal:stagg}
R. Portugal, R.A.M. Santos, T.D. Fernandes and D.N. Gonçalves, The Staggered Quantum Walk Model, Quantum Inf. Process. {\bf 15}, 85 (2016)

\bibitem{portugal:stagg:2}
R. Portugal, Staggered quantum walks on graphs, Phys. Rev. A {\bf 93}, 062335 (2016)

\bibitem{skw}
N. Shenvi, J. Kempe and K. B. Whaley, Quantum random-walk search algorithm, Phys. Rev. A {\bf 67}, 052307 (2003)

\bibitem{ambainis}
A. Ambainis, J. Kempe and A. Rivosh, Coins make quantum walks faster, in {\it Proceedings of the 16th ACM-SIAM Symposium on Discrete Algorithms}, 1099–1108 (2005)

\bibitem{potocek:search}
V. Poto\v cek, A. G\'abris, T. Kiss and I. Jex, Optimized quantum random-walk search algorithms, Phys. Rev. A {\bf 79}, 012325 (2009)

\bibitem{hein:search}
B. Hein and G. Tanner, Quantum search algorithms on the hypercube, J. Phys. A {\bf 42}, 085303 (2009)

\bibitem{reitzner:search}
D. Reitzner, M. Hillery, E. Feldman and V. Bu\v zek, Quantum searches on highly symmetric graphs, Phys. Rev. A {\bf 79}, 012323 (2009)

\bibitem{santos}
R. A. M. Santos, Szegedy's quantum walk with queries, Quantum Inf. Process., 1 (2016)

\bibitem{feldman:struct}
E. Feldman, M. Hillery, H. W. Lee, D. Reitzner, H. Zheng and V. Bu\v zek, Finding structural anomalies in graphs by means of quantum walks, Phys. Rev. A {\bf 82}, 040301(R) (2010).

\bibitem{hillery:probe}
M. Hillery, H. Zheng, E. Feldman, D. Reitzner and V. Bu\v zek, Quantum walks as a probe of structural anomalies in graphs, Phys. Rev. A {\bf 85}, 062325 (2012)

\bibitem{cottrell:find}
S. Cottrell and M. Hillery, Finding structural anomalies in star graphs: A general approach, Phys. Rev. Lett. {\bf 112}, 030501 (2014)

\bibitem{Lovett}
N. B. Lovett, S. Cooper, M. Everitt, M. Trevers and V. Kendon, Universal quantum computation using the discrete-time quantum walk, Phys. Rev. A {\bf 81}, 042330 (2010)

\bibitem{bose:pst}
S. Bose, Quantum communication through an unmodulated spin chain, Phys. Rev.
Lett. {\bf 91}, 207901 (2003)

\bibitem{wojcik:qw:pst}
P. Kurzynski and A. Wojcik, Discrete-time quantum walk approach to state transfer, Phys. Rev. A {\bf 83}, 062315 (2011)

\bibitem{gedik:qw:pst}
I. Yalcinkaya and Z. Gedik, Qubit state transfer via discrete-time quantum walks, J. Phys. A {\bf 48}, 225302 (2015)

\bibitem{zhan:qw:pst}
X. Zhan, H. Qin, Z. H. Bian, J. Li and P. Xue, Perfect state transfer and efficient quantum routing: A discrete-time quantum-walk approach, Phys. Rev. A {\bf 90}, 012331 (2014)

\bibitem{hein:wave:com}
B. Hein and G. Tanner, Wave Communication across Regular Lattices, Phys. Rev. Lett. {\bf 103}, 260501 (2009)

\bibitem{kendon:qw:pst}
V. M. Kendon and C. Tamon, Perfect State Transfer in Quantum Walks on Graphs, J. Comput. Theor. Nanosci. {\bf 8}, 422 (2011)

\bibitem{barr:pst}
K. Barr, T. Proctor, D. Allen and V. Kendon, Periodicity and perfect state transfer in quantum walks on variants of cycles, Quantum Inform. Comput. {\bf 14}, 417 (2014)

\bibitem{stef:pst}
M. \v Stefa\v n\'ak and S. Skoup\'y, Perfect state transfer by means of discrete-time quantum walk search algorithms on highly symmetric graphs, Phys. Rev. A {\bf 94}, 022301 (2016)

\bibitem{grover:search}
L. K. Grover, Quantum mechanics helps in searching for a needle in a haystack, Phys. Rev. Lett. {\bf 79}, 325 (1997)

\bibitem{krovi}
H. Krovi and T. A. Brun, Quantum walks on quotient graphs, Phys. Rev. A {\bf 75}, 062332 (2007)

\bibitem{note}
We note that the dimension of the invariant subspace can be reduced to three, however, the construction of the basis is more invovlved. Moreover, the subsequent analysis is not simpler than the one presented in the paper.



\end{thebibliography}
\end{document}